# 车路协同环境下交叉口交通控制仿真系统设计与实现

张蔚桐　　刘帅　　姚丹亚*

（清华大学 自动化系，北京，100084）

**摘要**　车路协同系统是智能交通系统的发展方向。目前车路协同系统实际部署较少，故相关研究仍以仿真实验为主要验证手段。传统商业交通仿真系统在宏观仿真方面更为出色，而对于小型交通场景较为冗余；亦不能实现车辆信息与车路协同路侧设备的信息交互。本文设计了车路协同环境下的交叉口交通控制仿真系统，提供丰富的用户设置参数，可以进行多种速率的仿真以满足不同的需求，为相关研究提供了测试基础。

**关键词**　车路协同；交叉口交通控制；仿真系统

## Design and Realization of Intersection Traffic Control Simulation System Using Connected Vehicle Technology

Zhang Weitong, Liu Shuai, Yao Danya*

*(Department of Automation, Tsinghua Univ., Beijing, 100084)*

**Abstract**　The connected vehicle technology is a remarkable trend in the field of intelligent transportation system. Since the actual deployment of the connected vehicle system is still lacking hitherto, simulation is widely adopted as the major method of verification in related researches. Although traditional commercial traffic simulation systems perform well in macroscopic simulations, they seem redundant in the circumstance of small-size traffic. In addition, those systems are unable to simulate the communication between vehicles and road-side infrastructures. This study designs a platform for the simulation of intersection traffic control using connected vehicle technology. By providing an abundance of customizable parameters, the platform can simulate at various levels of speed to meet difference requirements, serving as a basis for testing further research.

**keywords**　connected vehicle; intersection traffic control; simulation system

## 引言

车路协同系统将目前最先进的无线通信和互联网技术与传统的智能交通系统相结合，通过车与路侧设备和车与车之间的实时动态信息交互，实现交通的主动控制，提高道路通行效率，提升行车安全性。车路协同系统已成为智能交通领域的研究热点，是智能交通系统的发展方向[1-8]。

目前，车路协同技术的研究方兴未艾，示范系统的建设也已在多地开展，但尚无大规模部署。对车路协同环境下交通控制问题的研究，需要在正常交通流量时保证一定比例的装载设备的车辆才能与当前的交通控制形成有力的对比，体现出车路协同系统的应用价值。因此目前装车率的严重不足制约了交通主动控制在车路协同系统中的实地测

试，现有的车路协同环境下交通控制研究大多利用较为成熟的传统交通仿真系统进行仿真测试，从而对模型和算法进行评价。

对现有常用的交通仿真系统进行分析比对，可以发现将其应用于车路协同环境下交通控制问题的不足。故本研究设计了车路协同环境下的交叉口交通控制仿真系统。本文将在传统交通仿真系统分析基础上，从程序架构、软件界面、车辆生成算法和行驶策略等方面对这一自主研发的仿真系统进行介绍，并展示该仿真系统稳定性测试的结果。

# 1 传统交通仿真系统分析

国外针对交通仿真的研究取得了许多成果，已经形成了一些成熟的商用交通流仿真系统，如 VISSIM 仿真系统、PARAMICS（parallel microscopic simulator）仿真系统和 TransModeler 仿真系统等。相较于用于测试基本交通流模型的相对简单的程序，这些商用仿真系统有以下共同点[9, 10]：计算仿真能力强大，可以实现区域路网大量车辆的实时、半实时仿真；具有二维甚至三维动画效果，使得交通仿真结果更加直观；用户可以通过接口进行二次开发，以实现用户的仿真需求；具备交通信息和地图信息的输入系统，可以方便地输入交通路网和设施等信息；全面考虑了车辆出发地和目的地的路径选择，在不同线路上设定不同的交通需求压力；用户可以在一定程度上设计车辆跟驰和换道的行为；用户可以自主设计红绿灯信号配时；可以统计交通效率、安全性指标和环保指标。

以下对目前使用较为广泛的几款商业交通仿真软件进行简要的介绍。

## 1.1 VISSIM 仿真系统

VISSIM 仿真系统是 20 世纪 70 年代由德国卡尔斯鲁厄大学（University of Karlsruhe）进行早期设计的，随后由 PTV 公司开发成为了商用交通流仿真软件[11, 12]。VISSIM 仿真系统主要用于城市和郊区交通的模拟仿真中，是以时间为参照，以交通行为模型为基础的微观仿真系统。

VISSIM 仿真系统的跟驰模型建立在 Wiedemann 和 Reiter 提出的驾驶心理跟驰模型的基础上，但实际差别没有公开[13-17]。该模型将驾驶状态分为四种，即自由驾驶、接近、跟随和刹车。由于此模型也是本研究设计的仿真系统中车辆行驶的跟驰模型，详细模型描述在 4.3 节中给出，此处暂不展开说明。

## 1.2 Paramics 仿真系统

PARAMICS 是英国的 Quadstone PARAMICS 公司设计的交通仿真系统[18-20]。其一直是英国使用最多的微观仿真系统[21]。PARAMICS 主要用于城市中心区的各种主次路中，同样是以时间为参照，以交通行为模型为基础的微观仿真系统。早期 PARAMICS 采用了 Fritzsche 在 1994 年提出的驾驶心理跟驰模型[22, 23]。在其 5.0 版本以后的软件中采用了新的跟驰和换道模型，将车辆运动状态划分为六种：自由驾驶、接近、保持正常距离、加速保持距离、减速保持距离和危险刹车。

而 PARAMICS 的换道模型为始终采用了间隙接受模型。简单地说，如果车辆所在的车道和希望换到的车道的车辆都以恒定速度移动，则车辆希望在目标车道占据的位置前后都需要有不小于目标车距的间隙。而如果两个车道的速度不同，速度的差值可以折算成额外的间隙。

## 1.3 TransModeler 仿真系统

TransModeler 仿真系统由 MITSIM 仿真软件发展而来[24]，出自美国的 Caliper 公司，是该公司在 TransCAD 软件之后推出的一款多功能交通仿真软件。

TransModeler 的前身 MITSIM 仿真软件的跟驰模型是基于刺激-反应模型的，将驾驶过程划分为自由驾驶、跟随驾驶和紧急刹车三种状态。这一模型也在不断地进行改进，如 Yang 在其研究中的优化等[25, 26]。

MITSIM 模型中的换道行为一定程度上采纳了协同换道的思想，当希望换至的车道的后车准备接受车辆进入自己的车道时，会采用礼让函数的方式来预留一定的换道间隙[27-29]。

综合以上所述的仿真系统，它们在宏观仿真方面更为出色，而对于小型交通场景较

为冗余；控制策略的设定需要在其他平台进行二次开发并通过 COM 接口与仿真系统衔接；本文作者基于 VISSIM 系统进行的一些初步开发发现，即便不存在发生碰撞等事故的危险，一些内部机制的限制仍会使得设定的加速度控制信号不能完全被车辆使用。

而且，这些仿真系统只是给出了仿真中车辆的移动模型，并不能实现车辆信息与路侧智能单元的信息交互，更无法直接用以上仿真系统研究信息交互条件下交通控制的优化问题。在仿真过程中，仿真系统与用户的交互也不够充分。因此，本研究设计了一套仿真系统以更加真实地模拟车路协同环境下的交通控制。

## 2 车路协同环境下交叉口交通控制仿真系统架构

本研究设计的仿真系统使用 C++/Qt 进行开发，核心由一个 Traffic_v1 对象构成，这个对象的 UML（Unified Modeling Language，统一建模语言）图如图 1 所示。

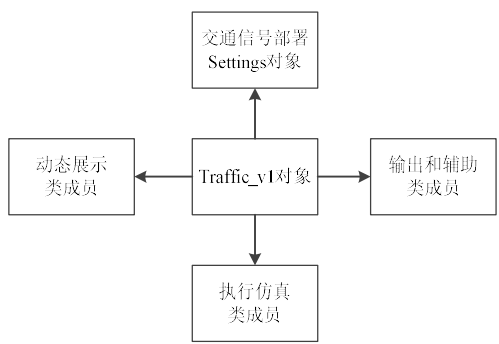

**图1 Traffic_v1对象的UML图**

仿真系统主要包括四个模块：显示模块、仿真模块、输出和辅助模块、信号配时模块。

### 2.1 显示模块

显示模块是仿真系统可视化显示的功能模块，其主要类成员如表 1 中所示。

程序每向前执行一步仿真就进行一次显示上的刷新，并将其绘制在交叉口中，整个函数完成如下步骤：

1) 绘制交叉口基本单元

根据信号配时的部署绘制红绿灯以及交叉口车道线。车道线显示长度为 50m，宽度为每条车道 3.5m，不考虑隔离带、路肩和非机动车道，双向六车道的宽度为 22.5m。

2) 绘制车辆

其中，开往交通路口的车辆可接受控制，而驶出路口的车辆按照自然驾驶策略继续行驶，自然驾驶的策略见 4.3 节。

**表 1 显示模块的类成员**

| 类成员 | 含义 |
| --- | --- |
| QPushButton* end | 停止按钮 |
| QPushButton* edit | 信号控制编辑 |
| QRadioButton* fast | 快速（100 倍速） |
| QRadioButton* medium | 中速（10 倍速） |
| QRadioButton* slow | 低速（1 倍速） |
| QRadioButton* very_slow | 极低速（0.1 倍速，仅用于 debug） |
| QLabel* speed | "speed" 标签 |
| QLabel* ratio | "Ratio Controlled" 标签 |
| QSlider* ratio_setting | 设置控制车辆比例的滑动条 |
| QLabel* ratio_shower | 显示控制车辆比例的标签 |
| com_label* st | 统计信息的类别的显示栏（12 个） |
| com_label* _st | 统计信息的数据显示栏（12 个） |
| QSlider* A_L | 设置车流量的滑块（西向，即 W 方向） |
| QSlider* B_L | 设置车流量的滑块（南向，即 S 方向） |
| QSlider* C_L | 设置车流量的滑块（东向，即 E 方向） |
| QSlider* D_L | 设置车流量的滑块（北向，即 N 方向） |
| QLabel*A_A | W 方向标签 |
| QLabel*B_A | S 方向标签 |
| QLabel*C_A | E 方向标签 |
| QLabel*D_A | N 方向标签 |
| QLabel*A_B | W 方向车流量数值显示 |
| QLabel*B_B | S 方向车流量数值显示 |
| QLabel*C_B | E 方向车流量数值显示 |
| QLabel*D_B | N 方向车流量数值显示 |

## 2.2 仿真模块

这是程序执行的核心模块，按照表 1 中 QRadioButton*设置的仿真速度（快速、中速、低速以及极低速），提供一个间隔为 10ms（快速模式），100ms（中速模式），1s（低速模式，即与现实世界的运行速率相同）或 10s（极低速，也是调试模式）的触发信号 QTimer*timer。接受时间触发信号之后，程序按照如图 2 所示的运行流程执行相应的操作。

在执行过程中，每个道路的入口将根据车辆生成算法来生成新的车辆。没有装载车路协同设备的车辆将按照自然驾驶策略行驶，而装载了的车辆将按照用户设定的行驶策略行驶。所有的用户策略均是调整车辆的加速度，在运动学仿真模块，平台将通过车辆现有的位置、速度和加速度计算车辆下一时刻的位置和速度。程序之后完成界面的重绘工作，并等待下一个时间触发信号的到来。最终，进程控制块根据用户设定的停止条件（在程序代码中设置好运行时间或在仿真界面上点击"END"键）终止仿真。

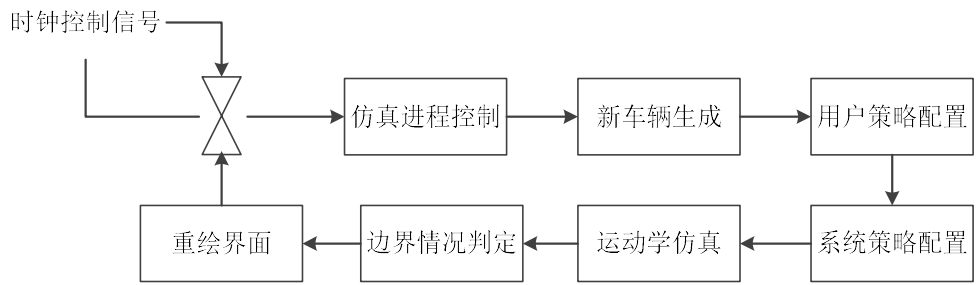

**图 2 仿真系统运行流程**

## 2.3 输出和辅助模块

输出辅助模块的主要工作在于将实时的仿真结果输出。随着仿真的进行，系统会建立四张.csv 格式的表格，存储在可执行文件（.exe）目录的根目录下的 result 文件夹中，result 文件夹中的每一个文件夹的命名规则为：

"日期+时间+仿真模式选择"

四张表分别记录了不同类型的数据：

1) car.csv

这张表记录了每辆车的初始速度（init_velocity）、理想到达时间（thoritical_time）、实际到达时间（act_time）以及后两者的差值（delta）。其中，理想到达时间为车辆以初始速度出发，以最大加速度加速到道路限速，之后保持匀速直线运动状态直到驶入交通路口内部所花费的时间；实际时间是车辆经过这段路程实际花费的时间；其差值即为车辆在这段路程中的延误。

2) stop.csv

这张表记录了每个时刻每个车道停车的总数和系统内停车的总次数，以及平均停车次数。

3) road.csv

这张表记录了每个时刻驶出某条车道的车辆总数和离开系统的车辆总数，以及其平均值。

4) stop_time.csv

这张表记录了每个时刻每个车道停车的总时间和系统内停车的总时间，以及平均每辆车的停车时间。

一般来说，road.csv 用来观测整个仿真过程是否正常，而其余三张表输出的是算法的评价指标结果。

其中，每当车辆的速度低于某一设定阈值（本文实验中设为 5km/h），即计为一次停车，停车时间随之不断累加，直至车辆速度超过阈值，而停车次数不随车辆的长期停车而改变。

## 2.4 信号配时模块

信号配时模块可实现用户自主设置信号周期和相位的需求。

在信号配时界面上，可以输入信号配时周期，并能通过移动时间线或输入时间线位

置，选中要更改的车道，点击三个按键"set red/green/yellow behind"中的某一个完成对时间线后面选中的车道的信号的设定。

## 3 系统界面和操作样例

本系统默认提供了一个双向六车道的路口，信号灯周期和相位可在仿真开始前预置。如图 3 所示，即为信号灯周期 90 秒，绿信比为 50%，每个相位的绿灯时间均为 45 秒。界面中，以"E""W""N""S"代指东（east）、西（west）、北（north）、南（south）四个来车方向，以"L""R""C"代指左转（left）、右转（right）、直行（center）三个车道，例如"WL"即代表从西向东的左转车道。

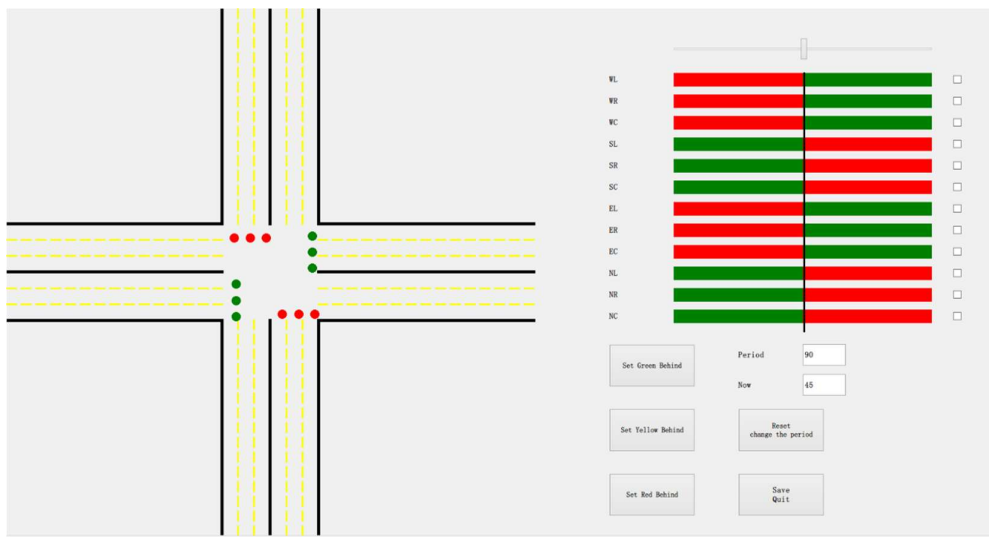

**图 3 车路协同环境下交叉口交通控制仿真系统信号预置界面**

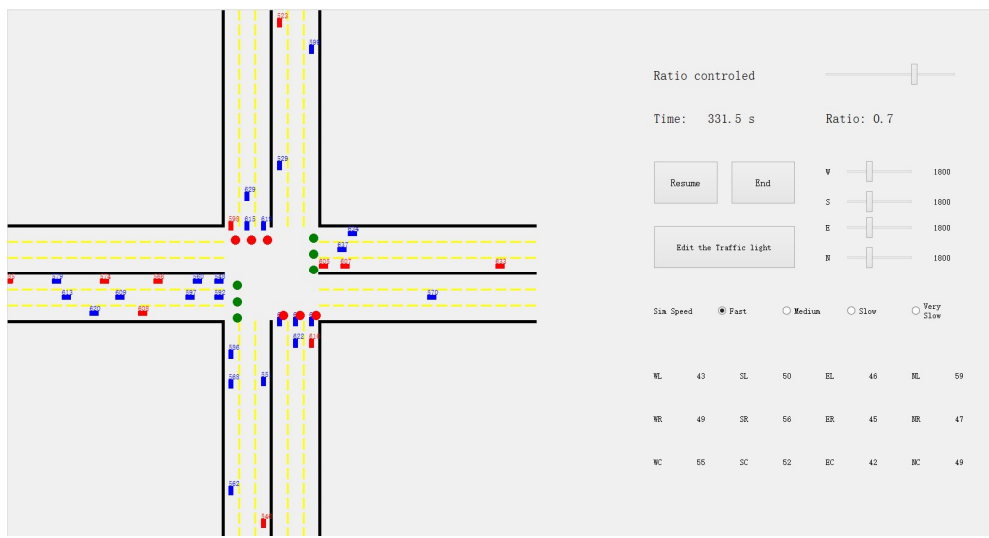

**图 4 车路协同环境下交叉口交通控制仿真系统运行界面**

在仿真开始前，使用者可以预置每个方向的流量和装有车路协同设备的车辆（即接受车速引导的车辆）的比例。仿真运行的速度可在界面上选择：快速（Fast）、中速（Medium）、低速（Slow）和极低速（Very Slow）。如 2.3.1 节所述，不同模式下，系统的时间触发信号间隔分别为 10ms（快速模式），100ms（中速模式），1s（低速模式）或 10s（极低速模式）。这就意味着，在"Fast"模式下，仿真系统运行 1 秒的仿真过程，相

574?1994-2018 China Academic Journal Electronic Publishing House. All rights reserved.　　http://www.cnki.net

当于实际道路中 100 秒的交通过程;"Slow"模式是与现实世界运行速率相同的模式;而"Very Slow"用来调试观察。图 4 所示,即为每个方向的流量是 1800 辆/小时,装车率为 70%,以"Fast"速度运行,已经运行了 331.5 秒的场景。其中蓝色为装有车路协同车载设备的车辆,而红色为没有安装车载设备的车辆。

# 4 车辆生成和行驶策略

## 4.1 程序基本参数和假定

首先,对车辆而言,车辆的加速度是可控的物理量,其他的主要物理量——速度和位置——可由车辆的加速度和当前时刻的速度根据运动学公式确定:

$$v(t+\Delta t) = v(t) + a(t)\Delta t \quad (1)$$

$$x(t+\Delta t) = x(t) + v(t)\Delta t + 0.5a(t)\Delta t^2 \quad (2)$$

由于车辆的加速度不能被准确地控制,仿真系统在策略给出加速度的准确值之后加入了一个正态分布的随机噪声。

同时,对车辆的加速度和速度也进行限制。车辆加速度绝对值的最大值被限制为 $a_{max}$,用以限制车辆的加速过程和刹车过程。另外,车辆的最大速度被限制为 $v_{max}$,车辆的最低速度被限制为 $v_{min}$。

不论选择了何种驾驶策略,如果策略给出的加速度值大于最大加速度 $a_{max}$ 或者小于 -$a_{max}$,抑或是计算得到的速度大于最大速度 $v_{max}$ 或小于最小速度 $v_{min}$,程序将自动将这些值设定为对应的边界值。如果车辆受到红灯影响而被迫刹车停车,最小速度限制将被设定为 0 来描述停车行为。但倒车行为不会出现在系统中。

为了简化模型,系统认为在所研究路段进行驾驶的过程中,车辆不进行变道行为。这个假设在现实情况中也是可以接受的,因为受到交通法规和规则的限制,车辆在交通路口前的变道行为往往是被禁止或限制的,车辆需要在路口停车线前一定距离完成相关的变道动作。

因此,在一个车道行进的车辆可以被描述成一个队列结构,即 $Q_{in}$。

## 4.2 车辆生成算法

系统假设车辆的到达过程是一个泊松过程,并假设泊松过程的强度是 $\lambda$,到达车辆的时间间隔 $S_n$, $S_{n-1}$ 可以被表述为

$$P(S_n - S_{n-1} \leq t) = 1 - e^{-\lambda t} \quad (3)$$

因为仿真过程的时间间隔被设定为 0.1s,每小时的平均车流量即为

$$\overline{F} = 36000\lambda \quad (4)$$

对于一个方向的车辆,车辆的小时流量被仿真界面上的交通流控制滑块设定之后,这个方向的车流量将按照特定的泊松过程进行生成 $\lambda = \dfrac{\overline{F}}{3600}$,并且被压入每个方向上的等待队列 $Q_{pending}$。

在生成车辆之后,需将其放入相应的车道,假设对于某个方向而言,不同车道来车是等概率分布的。此时需考虑车辆放入后能否安全行驶的问题。其具体定义为,车辆生成车头位置($S_{start}$)和该车道前一辆车的当前车头位置 $X_{-1}$ 之间的距离大于安全车头间距 $S_{safe}$,即

$$X_{-1} - S_{start} \geq S_{safe} \quad (5)$$

车辆生成时遍历得到所有符合上述安全条件的车道,并在其中等概率随机确定车辆放入的车道,车辆位置即为 $S_{start}$。

如果当前不存在合适的车道(适用于该方向车流量极大的情况),车辆将被保留在 $Q_{pending}$ 队列中并等待下一个仿真循环。

## 4.3 车辆跟驰策略

仿真系统采用了目前广泛使用的驾驶心理跟驰模型来描述不经控制的车辆运动行为。

对驾驶员心理反应的研究开始于 1960 年代。该研究的前提假设是:驾驶员在跟随前车驾驶的过程中,调整车速的主要依据是前后车的间距变化和速度变化,当这些变化超过阈值时,驾驶员感知到这种刺激并做出反应,调节车速[30-33]。

在这些假说中,视觉心理假说相对来讲最容易被人接受:在驾驶过程中,对前车进行跟随时驾驶员会根据前车后部在视野中

的面积大小以及通过视线来测量与前车的距离从而判断是靠近前车还是远离前车，驾驶员通过接受这样的一个刺激来判断与前车的行驶距离从而进行安全的操控[34]。

上个世纪 60 年代，研究发现了最小可认视角变化速率的存在，为视觉心理假设奠定了心理学基础[34]。Micheals 第一次提出了驾驶员心理跟驰模型的概念，是基于对驾驶员心理和生理的潜在因素进行分析而得出的结论[35]。他认为驾驶员是通过对出现在视野中前车的大小来判断前后车之间相对速度变化的。前车的大小具体是指前车在驾驶员视觉中投影夹角的变化。驾驶员通常是在车辆速度超过了感受视角变化的阈值时来选择加速或者减速。

一般认为感知阈值：

$$\frac{d}{dt}\left(\frac{\Delta v_{n,n-1}}{\Delta x_{n,n-1}^2}\right) \approx 6 \times 10^{-4} \quad (6)$$

在开放环境中的试验表明：在前车与后车距离接近时，驾驶员会在前车观察视角增大 10%时感觉到前后车距离发生了明显的变化；当前车与后车距离拉开时，驾驶员会在前车观察视角缩小 12%时感觉到前车与后车的距离发生了变化[33]。

在驾驶员不断接近或者远离前车的过程中阈值也逐渐发生变化。因此在 1970 年代人们更关注如何来确定平均阈值。Evans 和 Rothery 进行了一系列以知觉为基础的试验，尝试对 Micheals 提出的感知阈值进行量化[36,37]。试验要求在测试车内的乘客来判断与前车之间的距离是在变大还是在变小，并要求车内的乘客在设定的时间间隔（1s 或 2s）内观察目标前车同时做出距离判断。试验后的分析结果表明，判断车间距变化几率的正确函数可能是 $\Delta v_{n,n-1}$，$\Delta x_{n,n-1}$ 和观测时间的函数。

驾驶心理跟驰模型对驾驶状态的划分以及相应阈值的选取可以用 $\Delta v_{n,n-1}$—$\Delta x_{n,n-1}$ 相平面图来直观表达，如图 5 所示：

1) 自由驾驶状态：当与前车之间的距离大于最大相互作用距离时，驾驶员尽可能达到最大的行车速度，此时车辆处在自由驾驶的状态，也即图 5 中最上方的区域[34]。

2) 接近驾驶状态：当与前车之间的距离小于最大相互作用距离但大于刹车距离时，如果本车速度大于前车速度，则将不断接近前车，也即图 5 中右方扇形的区域。一般来说，在与前车逐渐接近的过程中，驾驶员会逐渐减速[34]。

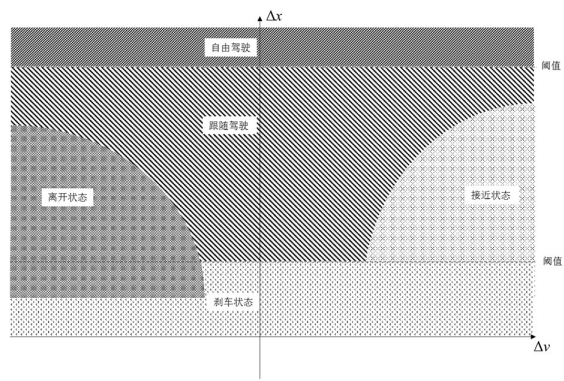

**图5 驾驶心理跟驰模型的阈值示意图**[34]

3) 离开驾驶状态：当与前车之间的距离小于最大相互作用距离但大于刹车距离时，如果本车速度小于前车速度，则将不断远离前车，也即图 5 中左方扇形的区域。当与前车之间的距离大于最大相互作用距离时，车辆从跟随驾驶状态转为自由驾驶状态[34]。

4) 跟随驾驶状态：当与前车之间的距离小于最大相互作用距离但大于刹车距离时，如果本车的速度与前车速度相近，则驾驶员会根据两车距离和前车速度来调整本车速度，车辆处于跟随驾驶状态，也即图 5 的中间区域[34]。

5) 刹车状态：当与前车之间的距离小于刹车距离时，出于安全的考虑，驾驶员将紧急制动刹车，也即图 5 中最下方的区域[34]。

当处在模型核心的跟随驾驶状态时，可以采用不同类型的模型进行描述，包括前面说的各种刺激-反应模型和安全距离跟驰模型。研究者同时也从驾驶员的生理和心理特性提出了一些新的模型。例如常被提及的 Leutzbach、Wiedemann 模型[37]：

$$a_n(t+T) = \frac{[\Delta v_{n,n-1}(t)]^2}{2[\Delta x_{n,n-1}(t) - S_{safe}]} + a_{n-1}(t) \quad (7)$$

其中 $S_{safe}$ 表示期望的最小安全跟随距离。

本研究采用的就是上述模型，其中车辆在自由驾驶状态下预期达到的最高车速平均采用道路限速的 90%。反应上限与 VISSIM 仿真系统相同，取为 150m，用以区分车辆是否处于自由驾驶状态。对于期望的最小安全跟随距离，根据《中华人民共和国道路交通安全法实施条例》[38]中关于安全车距的建议标准，本研究采用了与前车当前车速线性相关的模型来描述最小安全跟随距离：

$$S_{safe}(n,t) = \max\{\theta v(n-1,t), 5.5\} \quad (8)$$

式中 $\theta$ =0.55m×h/km 是线性因子，停车时的最小车头间距为 5.5m，安全车距取两者中的较大值。

### 4.4 路口排队的产生和消散

为了描述路口车辆排队的产生和消散过程，系统对每个车道引入准备到达路口的车辆队列 $Q_{in}$ 和停车等待队列 $Q_{block}$。

同 VISSIM 系统类似，定义反应上限 $S_{reaction}$，用以区分车辆是否处于自由驾驶状态，即当前后车距大于等于 $S_{reaction}$ 时，后车自由驾驶；前后车距小于 $S_{reaction}$ 时，后车将受到前车影响，采取跟驰等策略。

对于 $Q_{in}$，根据当前信号相位和队列的长度，采用的控制策略如下：

1) 当前信号为红灯

如果 $Q_{in}$ 中的第一辆车和 $Q_{block}$ 中的最后一辆车辆之间的距离小于 $S_{reaction}$ 时，队列 $Q_{in}$ 中的头车将刹车停至队列 $Q_{block}$ 最后一辆车之后。定义停车时前后两车的距离为 $S_{stop}$，则队列 $Q_{in}$ 头车的期望停车位置为队列 $Q_{block}$ 最后一辆车位置的 $S_{stop}$ 距离之后。

如果 $Q_{block}$ 为空并且 $Q_{in}$ 中的第一辆车和停车线之间的距离小于 $S_{reaction}$，$Q_{in}$ 中的头车以停车线 $S_{end}$ 为目标进行刹车。

如果 $Q_{in}$ 中的头车距离路口大于 $S_{reaction}$，$Q_{in}$ 中的头车将自由驾驶。

同时，$Q_{block}$ 中的车辆应当停车等待直到信号灯变绿。

2) 当前信号为绿灯

队列 $Q_{in}$ 中的第一辆车与队列 $Q_{block}$ 中的最后一辆车（如果存在）之间的车头距离不应小于安全车头间距 $S_{safe}$。如果两车之间车头距离小于安全车头间距，后车将采取制动行为。

综上所述，整个算法的简明思路可以用如下的伪代码表示：

//算法 1 准备到达路口的车辆队列 $Q_{in}$ 的控制策略

if (Q_block.empty())
    if (Light == Green) Q_in.first().drive_freely();
    else Q_in.first().brake_to(S_end);
else if (Q_block.last().pos-Q_in.first().pos < S_control)
    Q_in.first().brake_to (Q_block.last().pos-S_stop);
    else Q_in.first().drive_freely();

代码中 brake_to(desired_pos) 函数的意义是设置车辆的加速度使得车辆能够刹车到期望的位置上。

对于 $Q_{block}$，其产生和消散遵循如下的模型。

1) 产生/增长

当队列 $Q_{in}$ 中的第一辆车辆距离 $Q_{block}$ 中的最后一辆车辆（如果不存在，则设为停车线 $S_{end}$）的距离等于停车距离 $S_{stop}$ 时，这辆车将从 $Q_{in}$ 中移除并加入到 $Q_{block}$ 中。

2) 消散

当信号灯转绿灯时，队列 $Q_{block}$ 中的所有车辆将按照消散速度 $v_{dis}$ 向前移动，当车辆越过停车线时，将被从队列 $Q_{block}$ 中移除。

以上所述可以用下面的伪代码表示：

//算法 2 停车等待队列 $Q_{block}$ 的控制策略

if (light == green)
  Q_block.all_move_forward(v_dis)
else
  if(!Q_block.empty())
  if(Q_block.last().pos-Q_in.first().pos < S_stop)
    Vehicle v;
    v=Q_in.getfirst();
    Q_block.push(v);
  else if(S_end-Q_in.first().pos< S_stop)
    Vehicle v;
    v=Q_in.getfirst();
    Q_block.push(v);

## 5 仿真系统稳定性测试

为了测试仿真系统的稳定性，本研究进行了三个方面的具体测试。

### 5.1 长时间运行稳定性测试

进行十次 36000s（10 小时）长时间仿真，仿真过程中多次切换仿真速度，计算机保持正常使用（如编辑文档、打开网页等活动），如果系统内发生车辆碰撞、车道排队溢出（当交通流量大于饱和流量且运行较长时间将自然触发）等异常情况则仿真终止。实测未发生异常，可进行至少连续 36000s 的长时间仿真。

### 5.2 同时运行多个仿真稳定性测试

同时运行 3 个本仿真系统，系统中的信号周期、信号配时、车流量均不相同，进行十次 7200s（2 小时）的仿真，仿真过程中进行多次仿真速度间的切换，实测未发生异常，可同时运行多个仿真系统进行仿真。

### 5.3 仿真过程进入稳态测试

选取信号周期 90s，绿信比 50%，每个方向车流量 1800 辆/小时，进行 10800s（3 小时）的仿真，对车辆平均停车时间进行统计，统计结果如图 6 所示。

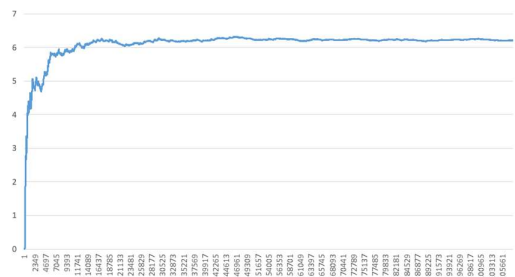

**图6 仿真系统进入稳态的过程图**

图中横坐标为采样点，每 0.1s 采样一次，故有 108000 条数据，纵轴是平均停车时间。由图中可以看出，大约在 1500 秒后运行的状况稳定下来，改变信号控制的设置、改变车流量或加入不同的车速控制算法都得到相似的结论。因此在实际仿真的过程中，用户可进行 7200 秒（2 小时）以上的长时间仿真，并去掉前 2000 秒的数据，以得到稳态的结果。

## 6 结语和展望

车路协同技术的发展正带来智能交通系统的变革。鉴于车路协同系统实际部署较少，仿真测试仍将是未来一段时间内学界研究车路协同环境下交通控制的重要方法。

传统交通仿真系统在宏观仿真方面更为出色，而对于单路口场景显得较为冗余；未能提供车辆与车辆、车辆与路侧设备交互的仿真环境；控制策略的设定需要在其他平台进行二次开发并通过 COM 接口与仿真系统衔接。

为了更有针对性地对车路协同环境下的交叉口的交通情况进行仿真，本文设计了车路协同环境下交叉口交通控制仿真系统，提供较多但不复杂的用户自主设置参量，可以进行高速和慢速仿真以满足不同的需求，而系统内车辆的自由行驶策略与 VISSIM 系统相似，采用的是 Leutzbach、Wiedemann 提出的心理跟驰模型。

系统稳定性测试结果表明，此仿真系统可以长时间、多个系统同时运行，且能够较快地进入稳态，具有较强的实用价值。

**参考文献**

[1] Willke T L, Tientrakool P, Maxemchuk N F. A survey of inter-vehicle communication protocols and their applications[J]. *IEEE Communications Surveys & Tutorials*, 2009, 11(2):3-20.

[2] Qu F, Wang F Y, Yang L. Intelligent transportation spaces: vehicles, traffic, communications, and beyond[J]. *IEEE Communications Magazine*, 2010, 48(11):136-142.

[3] Hartenstein H, Laberteaux L P. A tutorial survey on vehicular ad hoc networks[J]. *IEEE Communications magazine*, 2008, 46(6):164-171.

[4] Sukuvaara T, Nurmi P. Wireless traffic service platform for combined vehicle-to-vehicle and vehicle-to-infrastructure communications[J]. *IEEE Wireless Communications*, 2009, 16(6):54-61.

[5] Korkmaz G, Ekici E, Özgüner F. Supporting real-time traffic in multihop vehicle-to-infrastructure networks[J]. *Transportation Research Part C: Emerging Technologies*, 2010, 18(3):376-392.

[6] Bi Y, Cai L X, Shen X, et al. Efficient and reliable broadcast in intervehicle communication networks: A cross-layer approach[J]. *IEEE Transactions on Vehicular Technology*, 2010, 59(5):2404-2417.

[7] Luan T H, Ling X, Shen X S. Provisioning QoS controlled media access in vehicular to infrastructure communications[J]. *Ad Hoc Networks*, 2012, 10(2):231-242.

[8] Li L, Wen D, Yao D. A survey of traffic control with vehicular communications[J]. *IEEE Transactions on Intelligent Transportation Systems*, 2014, 15(1):425-432.

[9] Siegel J, Coeymans J E. An integrated framework for traffic analysis combining macroscopic and microscopic models[J]. *Transportation Planning and Technology*, 2005, 28(2):135-148.

[10] Hasan M, Jha M, Ben-Akiva M. Evaluation of ramp control algorithms using microscopic traffic simulation[J]. *Transportation Research Part C: Emerging Technologies*, 2002, 10(3):229-256.

[11] PTV The Transportation Experts. http://www.ptvamerica.com/vissim.html

[12] PTV GROUP. http://vision-traffic.ptvgroup.com/en-uk/home/

[13] Wiedemann R. Simulation des Straßenverkehrsflusses[R]. Institut für Verkehrswesen der Universität Karlsruhe, 1974.

[14] Weidemann R, Reiter U. Microscopic Traffic Simulation: The Simulation System-Mission[R]. Karlsruhe: University Karlsruhe, 1992.

[15] Wiedemann R, Reiter U. Microscopic traffic simulation: the simulation system MISSION, background and actual state[R]. In: CEC project ICARUS (V1052) Final Report. Brussels: CEC, 1992, 2:1-53.

[16] Reiter U. Empirical studies as basis for traffic flow models[C]// *Proceedings of the 2nd International Symposium on Highway Capacity*, 1994, 2:493-502.

[17] PTV AG Corporation. VISSIM 4.1 User Manual, 2005.

[18] SIAS Transport Planners, S-Paramics. http://www.sias.com/ng/sparamicshome

[19] Quadstone Paramics-Cutting Edge Microsimulation Software. http://www.paramics-online.com

[20] PARAMICS Programmer User Manual, version 6, Quadstone Limited, Edinburg, UK, 2007.

[21] S-Paramics 中国. http://www.paramics.cn/

[22] Fritzsche H T. A model for traffic simulation[J]. *Traffic Engineering and Control*, 1994, 35(5):317-21.

[23] Cameron G, Wylie B J N, McArthur D. Paramics: moving vehicles on the connection machine[C]// *Proceedings of the 1994 ACM/IEEE Conference on Supercomputing*. IEEE Computer Society Press, 1994:291-300.

[24] Caliper – Mapping Software, GIS and Transportation Planning Software. http://www.caliper.com

[25] Yang Q, Koutsopoulos H N. A microscopic traffic simulator for evaluation of dynamic traffic management systems[J]. *Transportation Research Part C: Emerging Technologies*, 1996, 4(3):113-129.

[26] Yang Q, Koutsopoulos H, Ben-Akiva M. Simulation laboratory for evaluating dynamic traffic management systems[J]. *Transportation Research Record: Journal of the Transportation Research Board*, 2000 (1710):122-130.

[27] Ahmed K I. Modeling drivers' acceleration and lane changing behavior[D]. Cambridge: Massachusetts Institute of Technology, 1999.

[28] Toledo T. Integrated driving behavior modeling[D]. Boston: Northeastern University, 2002.

[29] Choudhury C F. Modeling driving decisions with latent plans[D]. Cambridge: Massachusetts Institute of Technology, 2007.

[30] Todosiev E P. The action point model of the driver-vehicle system[D]. Columbus: The Ohio State University, 1963.

[31] Helander M. Applicability of drivers' electrodermal response to the design of the traffic environment[J]. *Journal of Applied Psychology*, 1978, 63(4):481-488.

[32] Boer E R. Car following from the driver's perspective[J]. *Transportation Research Part F: Traffic Psychology and Behaviou*r, 1999, 2(4):201-206.

[33] Ranney T A. Psychological factors that influence car-following and car-following model development[J]. *Transportation Research Part F: Traffic Psychology and Behaviour*, 1999, 2(4):213-219.

[34] 李力. 现代交通流理论与应用: 高速公路交通流.(卷 I)[M]. 北京: 清华大学出版社, 2011.

[35] Michaels R M. Perceptual factors in car following[C]// *Proceedings of the 2nd International Symposium on the Theory of Road Traffic Flow*. London: OECD, 1963.

[36] Evans L, Rothery R. Perceptual Thresholds in Car-Following—A Comparison of Recent Measurements with Earlier Results[J]. *Transportation Science*, 1977, 11(1):60-72.

[37] Leutzbach W, Wiedemann R. Development and applications of traffic simulation models at the Karlsruhe Institut für Verkehrswesen[J]. *Traffic Engineering & Control*, 1986, 27(5):270-278.

[38] 中华人民共和国国务院. 中华人民共和国道路交通安全法实施条例[Z]. 2004.04.30.